\documentclass[aps,prb,twocolumn,superscriptaddress,showpacs,floatfix]{revtex4}

\usepackage{graphicx}

\newcommand{\bra}{\langle}
\newcommand{\ket}{\rangle}
\newcommand{\mr}{{\mathbf{r}}}
\newcommand{\mR}{{\mathbf{R}}}
\newcommand{\mN}{{\mathbf{0}}}
\newcommand{\mk}{{\mathbf{k}}}

\newcommand{\ep}{\varepsilon}
\newcommand{\ph}{\varphi}
\begin{document}

\title{Unscreened Hartree-Fock calculations for metallic Fe, Co, Ni, and Cu 
from ab-initio Hamiltonians}

\author{I. Schnell}
\affiliation{Theoretical Division,
   Los Alamos National Laboratory, Los Alamos, New Mexico 87545}
\affiliation{Department of Physics, University of Bremen, P.O.Box 330 440,
   D-28334 Bremen, Germany}

\author{G. Czycholl}
\affiliation{Department of Physics, University of Bremen, P.O.Box 330 440,
   D-28334 Bremen, Germany}

\author{R. C. Albers}
\affiliation{Theoretical Division,
   Los Alamos National Laboratory, Los Alamos, New Mexico 87545}

\date{\today}

\begin{abstract}
Unscreened Hartree-Fock approximation (HFA) calculations for metallic
Fe, Co, Ni, and Cu are presented, by using a quantum-chemical approach. 
We believe that these are the first HFA results to have been done
for crystalline 3d transition metals.
Our approach uses a linearized muffin-tin orbital calculation to
determine Bloch functions for the Hartree one-particle
Hamiltonian, and from these obtains maximally localized Wannier
functions, using a method proposed by Marzari and Vanderbilt.
Within this Wannier basis all relevant one-particle and two-particle Coulomb
matrix elements are calculated.  The resulting second-quantized
multi-band Hamiltonian with ab-initio parameters is studied within the
simplest many-body approximation, namely the unscreened, self-consistent
HFA, which takes into account exact exchange and is free of self-interactions.
Although the d-bands sit considerably lower
within HFA than within the local (spin) density approximation L(S)DA, 
the exchange splitting and magnetic moments for ferromagnetic Fe, Co, and Ni
are only slightly larger in HFA than what is obtained either
experimentally or within LSDA.
The HFA total energies are lower than the corresponding LSDA calculations.
We believe that this same approach can be easily extended to include 
more sophisticated ab-initio many-body treatments of the
electronic structure of solids.
\end{abstract}

\pacs{71.10.Fd, 71.15.AP, 71.15.Mb, 71.20.Be ,71.45.Gm, 75.10.Lp}

\maketitle

\section{\label{sec:Introduction}Introduction}

In this paper we use a quantum-chemical approach
to present unscreened Hartree-Fock approximation (HFA)
calculations of metallic Fe, Co, Ni, and Cu. Because
our approach uses localized Wannier functions, it
is a Hubbard-like method that should be
easily generalized to include more sophisticated many-body
treatments of correlation effects.  Nonetheless, it is useful to understand
what a HFA method would give before moving on to consider correlation.
To place these calculations in context it is useful to briefly
review the status of electronic-structure calculations in solids.

Most existing ab-initio (first-principles) methods for the numerical
calculation of the electronic properties of solids are based on density
functional theory (DFT)\cite{HK64}, which in principle is exact and
properly takes into account many-body effects involving the
Coulomb interaction between the electrons; for an overview on the
present status of DFT we refer to the books\cite{DrGross90,Eschrig96}.
But, in general, the functional dependence of the kinetic energy and the
exchange and correlation part of the Coulomb (interaction) energy on the
electron density are not known explicitly, and hence additional
approximations and assumptions are necessary.  A well established
additional approximation is the local density approximation
(LDA)\cite{KS65} (or local spin-density approximation, LSDA, for
magnetic systems), which assumes that the exchange-correlation potential
depends only on the electronic density locally.
Even then, the functional dependence of the
exchange-correlation energy on the density is not known in general, and
it is usually necessary to make an ansatz for the exchange-correlation
functional, which is based on the homogeneous electron gas.
The LDA goes beyond the simplest electron-gas approximation, the HFA,
in that correlation energy is explicitly taken into account.  On the
other hand, the exact HFA exchange potential is non-local, an effect which the
local LDA exchange potential misses.
However, in practice LDA-treatments are
simpler than HFA-calculations, because local exchange is easier to treat than
non-local exchange, and are usually in better agreement with experiment.
Therefore, DFT-(LDA-like) treatments have been far more common than HFA
during the past few decades, even in quantum chemistry (with a long
tradition of methods based on HFA).

Ab-initio DFT-LDA calculations have been very successful
for many materials and ground-state properties such as crystal structure,
ground state and ionization energy, lattice constant, bulk modulus, crystal
anharmonicity\cite{Rose84}, magnetic moments, and some photo emission
spectra. However, there are also important limitations.
For example, LDA predicts a band gap for semiconductors that is almost a
factor of two too small, while the HFA overestimates the band gap for
semiconductors\cite{Svane86}.
In addition, for many strongly correlated
(narrow energy band) systems such as high-temperature superconductors,
heavy fermion materials, transition-metal oxides, and 3d itinerant
magnets, the LDA is usually not sufficient for an accurate description
(predicting metallic rather than semiconducting behavior, failing to
predict quasi-atomic-like satellites, etc).

Therefore, it is important to look for ab-initio methods and improvements that
go beyond L(S)DA.  Recently there have been several attempts to combine
ab-initio LDA calculations with many-body
approximations.\cite{AZA91,SASS91,SAS9294,APKAK97,LK98,DJK99,KL99,LL00,
WPN00,NHBPAV00} All of these recent developments add local, screened
Coulomb (Hubbard) energies U between localized orbitals to the
one-particle part of the Hamiltonian obtained from an ab-initio LDA
band-structure calculation, but differ in how they handle the
correlation part.  These approaches, have in common that they have to
introduce a Hubbard U as an additional parameter and hence are not really
first-principles treatments.  Although they use an LDA ab-initio method to
obtain a realistic band structure, i.e., single-particle properties, Coulomb
matrix elements for any particular material are not known, and the
Hubbard U remains an adjustable parameter.  In addition, since some
correlations are included in LDA as well as by the Hubbard U, it
is unclear how to separate the two effects and double-counting of
correlation may be included in these approximations.

Other attempts to improve LDA include gradient corrections, non-local density
schemes, self-interaction corrections, and the GW approximation (GWA).
Gradient corrections\cite{DrGrossKap7} approximately account for the fact that
the electron density is not constant but $\mr$-dependent in an inhomogeneous
electron gas and use an exchange-correlation potential containing $\nabla
n(\mr)$ terms.  The non-local density schemes go beyond LDA by
considering that the exact exchange-correlation potential
$V_{\text{xc}}(\mr)$ cannot depend only on the density $n(\mr)$ at the
same position $\mr$ but should depend also on the electron density at
all other positions $n(\mr')$.  Usually the new ansatz for the functional
of the exchange-correlation energy contains the pair correlation
function or the interaction of the electrons with the
exchange-correlation hole.\cite{Gunnarsson80,DrGrossKap7}.  The recently
developed exact exchange (EXX) formalism\cite{Vogl97,Kotani97} cancels the
spurious (unphysical) electronic self-interaction present in LDA and
gradient corrected exchange functionals.  A standard method for ab-initio
calculations of excited states is the GWA\cite{Hedin,Aulbur}.
Denoting the one-particle Green function by
$G$ and the screened interaction by $W$, the GWA is an approximation for
the electronic selfenergy $\Sigma \approx GW$, which is correct in
linear order in $W$ and can diagrammatically be represented by the
lowest-order exchange (Fock) diagram.
The one-particle Green function $G$ is usually obtained for the effective
one-particle LDA Hamiltonian.

The HFA has long been a standard electronic-structure method.
Despite its many manifest defects, it 
is still important to know what such a calculation would predict 
before turning to more sophisticated approaches for correlation effects.
In this paper we provide HFA calculations for Fe, Co, Ni, 
and Cu using an approach that we hope will be easily generalizable to 
more sophisticated treatments of correlation.

This is done by using the following steps:
\begin{enumerate}
\item Perform a conventional, self-consistent, band-structure
   calculation for an effective one-particle Hamiltonian, namely, the
   Hartree Hamiltonian, to obtain a suitable basis set of Bloch
   functions.
\item By taking into account only a finite number $J$ of bands one
   chooses a truncated one-particle Hilbert space. The
   Marzari-Vanderbilt\cite{MV97} algorithm is then used to construct a
   maximally localized set of Wannier functions, which span the same
   truncated one-particle Hilbert space.
\item All one-particle (tight-binding) and two-particle (Coulomb)
   matrix elements of the Hamiltonian within this Wannier function basis
   are calculated.
\item The resulting electronic many-body Hamiltonian in second
   quantization with parameters determined from first principles is
   studied within the HFA.
\end{enumerate}
We use the ``linear muffin-tin orbital'' (LMTO) method\cite{OKA75} within the
``atomic-sphere approximation'' (ASA)\cite{Skriver} to
perform the band-structure calculation for the Hartree Hamiltonian in
first quantization.
The second step of constructing localized Wannier functions
is important, since then both the tight-binding and Coulomb
matrix elements should be important only on-site and for a few neighbor
shells (the most natural mapping to standard Hubbard-like models).
The direct Coulomb matrix elements of the
maximally localized Wannier basis are rather large, about 20 eV in
magnitude.  Our results are compared with those
obtained from a standard LSDA calculation.\cite{DrGross90,Eschrig96,MJW78,Sch02}
Although the 3d-bands and the 4s-band overlap in the L(S)DA
approximation, our unscreened HFA calculations give 3d-bands that lie
considerably lower (between 10 and 20 eV) than the 4s-band.  The HFA
correctly predicts ferromagnetism for the ferromagnetic metals Fe, Co,
and Ni and no magnetism for Cu, but with a much larger exchange
splitting between majority and minority 3d bands than obtained within
LSDA and with a slightly larger magnetic moment per site than obtained
experimentally or within LSDA.  The total energy is lower in HFA than in LSDA.
The LSDA results for metals are probably more reliable than our
HFA results, which lack important screening and correlation effects.
In order for our method to go beyond LSDA we would need
to use better many body methods than the (unscreened) HFA, which
should be possible within our scheme.

To the best of our knowledge we do not know of any other published HFA
results (band structure, density of states, magnetism, magnetic
moment, total energy, etc.) for the 3d ferromagnets Fe, Co and
Ni, unless it was implicitly applied to these materials for schemes
like the local ansatz\cite{Stollhoff}, where HFA results serve as an
input to higher order calculations.
This is not surprising since the HFA has, from very early on, been
viewed as a poor approximation for metals.
For example, when applied to
the homogeneous electron gas (as the simplest model of an infinite
metallic system), the HFA has well-known Fermi edge
singularities\cite{Mahan,Pisani}. These lead, in particular, to a
vanishing density of states (DOS) at the Fermi energy, which is, of
course, unphysical.  This unphysical feature usually prevails in actual
HFA-calculations for real metals\cite{Monkhorst79}, though sometimes
this singularity is hard to see in actual
HFA-results\cite{DovesiPisani82}.
In our calculations the
non-locality is handled through the calculation of expectation values
(matrix elements of the density matrix), which makes HF calculations
as easy as Hartree calculations.
Furthermore, because of our localized Wannier basis,
we only keep on-site and next neighbor Coulomb and
exchange matrix elements.  Hence our calculations have an effective
short-ranged Coulomb interaction.  Although longer-range Coulomb
matrix elements are small in our calculations, which is why we
truncate them, it is possible that if all of them were kept to
infinite distances that they could add up to give Fermi edge
singularities (which are due to the long-ranged nature of the bare
Coulomb interaction) and other standard anomalies.  Correlation or
screening would quickly kill these effects.

The approximation closest to HF is the exact exchange formalism
(EXX)\cite{Vogl97,Kotani97} mentioned earlier.
The EXX method is different from the LDA only in that the EXX
energy\cite{Kotani97}, rather than the LDA exchange energy, is used;
thus, LDA correlations are still present.
The EXX energy, which corresponds to the Fock term in the HFA,
is treated as a functional of electron density and the method
is also (like HF) self-interaction-free.  Although the EXX would appear
to be very similar to HF, the EXX-only method\cite{Kotani94,KotaniPRL95},
which does not include any correlation, gives the dispersion of noninteracting
electrons instead of the HF dispersion when applied to the homogeneous
electron gas, while their total energies are exactly the same.
For atoms the EXX-only method gives total energies that agree well with HF.
Due to these similarities, we will compare our results with EXX where possible.
One should note, however, that most EXX calculations include a local
correlation potential.

The paper is organized as follows. In Section \ref{sec:Ham}, we
briefly summarize some basic notation, give the Hamiltonian in first and
second quantization, and describe our LMTO-Hartree calculations and our
Wannier basis functions.
Results for the matrix elements, in particular, the direct Coulomb and exchange
matrix elements are given in Section \ref{sec:matrixel}; we also
compare these results with calculations of the Slater integrals.
The application of the (unscreened) HFA to the multi-band
Hamiltonian in second quantization is the subject of Section
\ref{sec:HartreeFock}. For an interpretation of the results we compare the
numerical HFA results obtained for the crystal with previous
atomic HFA results and with numerical and analytical results for a simplified
local atomic model in Section \ref{sec:atomicHFA}.
A comparison with the more standard LSDA as well as EXX results follows in
Section \ref{sec:LSDA}, before the paper closes with a short discussion.

\section{\label{sec:Ham}Hamiltonian and Basis Functions}

A system of $N_e$ interacting (non-relativistic) electrons can be
described by the Hamiltonian
\begin{equation} \label{eq:firstquant}
   H = T + V + W =
   \sum_{i=1}^{N_e} \frac{\mathbf{p}_i^2}{2m} +
   \sum_{i=1}^{N_e} V(\mr_i) +
   \sum_{i > j} \frac{e^2}{|\mr_i-\mr_j|} ~~.
\end{equation}
The first part $T$ is the kinetic energy of the electrons. The $V(\mr)$
describes the external one-particle potential.
The formalism of ``second quantization'', automatically accounts for the
antisymmetry through the fermion anticommutation relations. In second
quantization the full many-body Hamiltonian can be written as:
\begin{equation} \label{eq:secondquant}
   H = \sum_{i,j,\sigma} t_{ij} c_{i\sigma}^{\dagger} c_{j\sigma} +
   \frac{1}{2} \sum_{i,j,k,l,\sigma,\sigma'}
   W_{ij,kl} c_{i\sigma}^{\dagger}c_{j\sigma'}^{\dagger}c_{k\sigma'}c_{l\sigma}
\end{equation}
Here $i,j,k,l$ denote a complete set of one-particle orbital quantum
numbers and $\sigma,\sigma'$ are the spin quantum numbers.  The states
$|i\ket$ and the corresponding wave functions $\ph_i(\mr)=\bra\mr|i\ket$
form a basis of the one-particle Hilbert space.  The matrix elements in
Eq. \ref{eq:secondquant} of course depend on the one-particle basis
$\{|i\ket\}$ that is chosen.
But because of the completeness relation the physical results should,
not depend on the choice of the one-particle basis.
Because of the lattice
periodicity an obvious choice for a one-particle basis is a Bloch basis
$\{|n\mk\ket\}$; then the orbital one-particle quantum numbers $n,\mk$
are the band index $n$ and the wave number $\mk$ (within the first
Brillouin zone).
In practice, one can work only on a truncated, finite-dimensional
one-particle Hilbert space. Here the truncation consists of including
only a finite number of bands and a set of
$\mk$-values from a discrete mesh in $\mk$-space.
But, because the Bloch states are delocalized, a very large number of Coulomb
matrix elements (depending on four quantum numbers) between
all possible $\mk$-states would have to be calculated.
Therefore, it seems that a more
appropriate basis would be to use well localized wave functions,
where it is expected that a short-range tight-binding assumption will
hold, i.e., that the on-site and the inter-site matrix elements for only
a few neighbor shells are sufficient.
The Wannier states are related to the Bloch states by the unitary
transformations:
\begin{eqnarray} \label{eq:wannier1}
   w_{\mR n}(\mr) &=& \bra \mr|\mR n\ket
   = \frac{1}{N} \sum_{\mk} e^{-i\mk\mR}\psi_{n\mk}(\mr)
\\ \nonumber
   |\psi_{n\mk}\ket &=& \sum_{\mR} e^{i\mk\mR} |\mR n\ket
\end{eqnarray}
Now our strategy is the following:
\begin{itemize}
\item Perform a traditional band-structure calculation for an effective
   one-particle Hamiltonian $H_{\text{eff}}$ with lattice periodicity
   to obtain a Bloch basis of the Hilbert space.
   Only a finite number of band indices will be considered and
   the calculations will be done for a discretized,
   finite mesh in $\mk$-space, i.e., we will work only on a
   reduced, truncated Hilbert space.
\item Determine well-localized Wannier functions spanning the same
   (truncated) Hilbert space as the Bloch basis from the
   canonical transformation (\ref{eq:wannier1}) described above.
\end{itemize}
Of course, the important energy bands (and corresponding band indices)
are those that determine the electronic properties of the system, i.e.,
the bands near to the Fermi level. Because the Hilbert space is
truncated, we do no longer work with a complete basis set. Hence, it is
important to start from Bloch functions obtained from a band-structure
calculation for a well chosen effective one-particle Hamiltonian.
The simplest choice would be the bare one-particle potential $V(\mr)$.
However, without any Coulomb repulsion the
3d-states become very strongly bound atomic-like (core) states, which
would be pushed well below the Fermi energy, and therefore the
corresponding Bloch eigenfunctions are not a good starting point to
describe the electronic bands close to the Fermi level.
Because the Hilbert space is truncated,
it is extremely important to start from a
band Hamiltonian $T+V_{\text{eff}}$ that gives eigenfunctions as close as
possible to those which are expected to form the relevant many-body states of
the interacting system.
The bare one-particle potential is consequently a bad choice.
Therefore, we choose the Hartree Hamiltonian, which
already accounts for effects of the Coulomb interaction in the mean-field
approximation.  Therefore,
the eigenenergies (energy bands) will be about the right magnitude and
the resulting basis functions can be expected to be more suitable in the
energy regime around the Fermi level.
Then the Bloch basis is obtained
by solving the one-particle Schr{\"o}dinger equation
\begin{equation} \label{eq:Hartree1quant}
   \Big(\frac{{\bf p}^2}{2m} + V(\mr) +
   V_{\text{H}}(\mr)\Big)\psi_{n\mk}(\mr)
   = \ep_{n}(\mk) \psi_{n\mk}(\mr)
\end{equation}
where the Hartree potential is given by
\begin{equation} \label{eq:Hartreepotential}
   V_{\text{H}}(\mr) = \int d^3r' \frac{e^2 \rho(\mr')}{|\mr - \mr'|} ~~.
\end{equation}
Since the only purpose in solving the effective one-particle
Schr{\"o}dinger equation (\ref{eq:Hartree1quant}) is the construction of
a suitable basis set of Bloch functions, we will not make use of the
eigenenergies $\ep_{n}(\mk)$ obtained in Eq. \ref{eq:Hartree1quant}.
Note that the Hartree potential, and hence our basis,
is independent of spin.  Nevertheless we can (in the following) expand the
spin dependent HF Hamiltonian in this basis.

For the materials of interest
we performed a self-consistent Hartree band-structure calculation.
Besides the nuclear charge we used the (experimentally) known results for
the lattice structure (bcc for Fe, fcc otherwise; Co should actually
be hexagonal) and for the lattice constant as input.
For the band-structure calculation we used the LMTO-ASA
method\cite{Skriver,OKA75} within the atomic sphere approximation (ASA).
We have used the frozen core approximation\cite{Skriver}, i.e. only
treated the valence electrons as actual bands,
while leaving the core electrons ``frozen''.
For the radius of the overlapping muffin-tin spheres, the Wigner-Seitz
radius $S$, we used: $S=2.662 a_0$ for Fe, $S=2.621 a_0$ for Co, $S=2.602 a_0$
for Ni and $S=2.669 a_0$ for Cu (Ref.~\onlinecite{Skriver}).
Within the muffin-tin spheres the
potential and wave functions are expanded in spherical harmonics with
a cutoff $l_{\rm max}=2$, i.e., s, p, and d-orbitals are included,
which limits the calculation to 9 bands for one atom per unit cell.

In Ref. \onlinecite{SCA02}, we describe how maximally localized Wannier
functions can be calculated from LMTO Bloch wave functions using a
method proposed by Marzari and Vanderbilt, which is
described in detail in Ref. \onlinecite{MV97}.
The Wannier functions are
admixtures having different angular contributions (3d, 4s, 4p).
Since the original Bloch functions from which
the Wannier functions are constructed
were given in terms of a spherical harmonics expansion, the new Wannier
functions (and their contribution in each individual muffin-tin sphere)
can also be decomposed into these spherical harmonics contributions
\begin{equation} \label{eq:wann-wave}
   w_n(\mR;\mr) = \sum_L \left\{
     \phi_{\nu l}(r) A_L^{\mR n} + \dot \phi_{\nu l}(r) B_L^{\mR n}
   \right\}
   Y_L( \hat \mr ) ~~.
\end{equation}
One can then calculate the weight of the contributions to the Wannier
function (centered at $\mN$) within the different muffin-tin spheres
\begin{equation} \label{eq:wnR}
   \bra w_n | w_n \ket_\mR \equiv \int_\mR d^3\mr |w_n(\mr)|^2
   = \int_\mN d^3\mr |w_n(\mR;\mr)|^2 ~~,
\end{equation}
and one can also decompose this into the different $l$-contributions
according to:
\begin{equation} \label{eq:wnR_eval}
   \bra w_n | w_n \ket_\mR = \sum_l
   \underbrace{
     \sum_{m=-l}^l
     \left\{
       |A_{l m}^{\mR n}|^2 + \bra \dot\phi_{\nu l}^2 \ket |B_{ l m}^{\mR n}|^2
     \right\}
   }_{ \equiv ~ C_l^{\mR n} }
\end{equation}
For the 3d-system iron these quantities are tabulated in Table
\ref{tab:fe_wf}. The first line is the weight $\bra w_n|w_n \ket_{\mN}$
in the center muffin-tin. Between 88 and 98\%
of the total weight of the Wannier functions is to be found already within
the center muffin-tin; this shows how well localized our Wannier functions
are with the lowest five functions having values of more than 95\%.
Rows 2--4
in this table indicate the different $l$-contribution or $l$-character of the
Wannier functions. One sees that the optimally localized Wannier
functions are not pure within their $l$-character, but the
lowest five Wannier functions (0-4) still have mainly $l=2$ (3d) character.
Higher band-index states (which are slightly less well localized according to
row 1) are admixtures that have mainly $l=1$ (4p) character (about 50 \%),
but also a considerable amount of $l=0$ (4s) and $l=2$ (3d) character.
Corresponding results for the other 3d-systems Co, Ni, and Cu are similar
and, therefore, not repeated here.
Our detailed results are given in Ref. \onlinecite{Sch02}.

% tab:fe_wf
\begin{table}[!ht]
   \renewcommand{\arraystretch}{1.2}
   \begin{tabular}{c|ccccccccc}
     $n$ & 0 & 1 & 2 & 3 & 4 & 5 & 6 & 7 \\
   \hline
$\sum_l C_l^{\mN n}$ & .9761 & .9765 & .9596 & .9800 & .9773 & .8754 
& .8731 & .8763 \\
$\sum_\mR C_{l=0}^{\mR n}$ & .0019 & .0018 & .0081 & .0019 & .0017 & 
.2224 & .2381 & .2265 \\
$\sum_\mR C_{l=1}^{\mR n}$ & .0955 & .0726 & .1797 & .0611 & .0728 & 
.5480 & .5509 & .5347 \\
$\sum_\mR C_{l=2}^{\mR n}$ & .9026 & .9256 & .8121 & .9370 & .9255 & 
.2295 & .2110 & .2388 \\
   \end{tabular}
   \caption{\label{tab:fe_wf} Some properties of the lowest eight maximally
     localized Wannier functions of Fe.}
\end{table}

\section{\label{sec:matrixel}One particle and Coulomb matrix elements}

 From the optimally localized Wannier functions we calculate the
one-particle matrix elements
\begin{equation} \label{eq:t12}
   t_{12} = \int d^3\mr~ w_1^*(\mr)\left(-\frac{\hbar^2}{2m}\nabla^2 +
   V(\mr)\right)w_2(\mr)
\end{equation}
and the Coulomb matrix elements of the Hamiltonian
\begin{equation} \label{eq:W1234}
   W_{12,34} = \int d^3\mr~ d^3\mr'~ w^*_1(\mr) ~ w^*_2(\mr') ~~
   \frac{e^2}{|\mr-\mr'|} ~~ w_3(\mr') ~ w_4(\mr) ~~.
\end{equation}
Here we use the abbreviated notation 1 to mean
$\mR_1n_1$ and 2 to mean for $\mR_2n_2$, etc.
In Ref. \onlinecite{SCA02}, we have described how these matrix elements
can be evaluated.  Concerning the Coulomb matrix elements,
we have used the two different numerical algorithms proposed in
Ref. \onlinecite{SCA02} for their evaluation, namely the FFT-algorithm
and a spherical expansion algorithm.
The latter method makes use of the fact that (in each muffin-tin sphere)
the Wannier functions are explicitly given as linear combinations of
products of spherical harmonics and a radial wave function.
The expansion
\begin{equation} \label{eq:p37}
   \frac{1}{|\mr-\mr'|} = \sum_{k=0}^\infty
   \frac{4\pi}{2k+1} ~ \frac{r_<^k}{r_>^{k+1}} \sum_{m=-k}^k
   Y_K^*(\hat\mr') ~ Y_K(\hat\mr)
\end{equation}
($K=\{k,m\}$) makes it possible to express the on-site Coulomb integrals as
one-dimensional integrals over products of the radial functions and
Gaunt coefficients. The results obtained by this algorithm
and by the independent FFT-algorithm agree within errors
of at most 1\%.
Since our basis functions are well-localized, we may truncate the
tight-binding and Coulomb matrix elements.
We only consider on-site and next neighbor matrix elements,
by next neighbor Coulomb matrix elements, we mean matrix elements
for which the four sites (appearing in the indices) are (pairwise) maximally
a next neighbor distance apart.

% tab:fe_UJ
\begin{table}[!ht]
   \renewcommand{\arraystretch}{1.2}
   \begin{tabular}{c|ccccccccc}
      $U_{nm}$ & 0 & 1 & 2 & 3 & 4 & 5 & 6 & 7 & 8 \\
   \hline
0 & 22.42 & 20.90 & 20.10 & 20.96 & 20.86 & 14.16 & 13.32 & 13.96 & 13.50 \\
1 & 20.90 & 23.04 & 19.95 & 21.55 & 21.53 & 14.07 & 13.54 & 13.58 & 14.15 \\
2 & 20.10 & 19.95 & 20.77 & 20.05 & 19.83 & 12.95 & 13.46 & 13.37 & 13.22 \\
3 & 20.96 & 21.55 & 20.05 & 23.27 & 21.67 & 13.46 & 14.05 & 13.98 & 13.98 \\
4 & 20.86 & 21.53 & 19.83 & 21.67 & 22.99 & 13.71 & 13.28 & 14.25 & 14.12 \\
5 & 14.16 & 14.07 & 12.95 & 13.46 & 13.71 & 13.67 &  9.45 &  9.58 &  9.64 \\
6 & 13.32 & 13.54 & 13.46 & 14.05 & 13.28 &  9.45 & 13.52 &  9.27 &  9.50 \\
7 & 13.96 & 13.58 & 13.37 & 13.98 & 14.25 &  9.58 &  9.27 & 13.75 &  9.65 \\
8 & 13.50 & 14.15 & 13.22 & 13.98 & 14.12 &  9.64 &  9.50 &  9.65 & 13.81 \\
   \end{tabular}
   \vspace{5mm}
   \quad
   \begin{tabular}{c|ccccccccc}
      $J_{nm}$ & 0 & 1 & 2 & 3 & 4 & 5 & 6 & 7 & 8 \\
   \hline
0 & 22.42 &  0.84 &  0.61 &  0.75 &  0.99 &  0.86 &  0.73 &  0.81 &  0.42 \\
1 &  0.84 & 23.04 &  0.77 &  0.88 &  0.84 &  0.70 &  0.51 &  0.48 &  0.86 \\
2 &  0.61 &  0.77 & 20.77 &  0.88 &  0.70 &  0.96 &  0.93 &  0.92 &  0.60 \\
3 &  0.75 &  0.88 &  0.88 & 23.27 &  0.82 &  0.33 &  0.78 &  0.64 &  0.69 \\
4 &  0.99 &  0.84 &  0.70 &  0.82 & 22.99 &  0.52 &  0.46 &  0.75 &  0.83 \\
5 &  0.86 &  0.70 &  0.96 &  0.33 &  0.52 & 13.67 &  0.58 &  0.56 &  0.57 \\
6 &  0.73 &  0.51 &  0.93 &  0.78 &  0.46 &  0.58 & 13.52 &  0.45 &  0.56 \\
7 &  0.81 &  0.48 &  0.92 &  0.64 &  0.75 &  0.56 &  0.45 & 13.75 &  0.55 \\
8 &  0.42 &  0.86 &  0.60 &  0.69 &  0.83 &  0.57 &  0.56 &  0.55 & 13.81 \\
   \end{tabular}
   \caption{\label{tab:fe_UJ} On-site direct and exchange
     Coulomb matrix elements between Wannier functions for Fe.
     All energies are in eV's.}
\end{table}

Results for the on-site direct and exchange Coulomb
matrix elements between the optimally localized Wannier functions are
given in Table \ref{tab:fe_UJ} for iron.
The direct Coulomb integrals $U_{nm} = W_{nm,mn}$ between the Wannier states
with the lowest five band indices ($n,m \in \{0,\ldots,4\}$),
which according to the Table \ref{tab:fe_wf}
have mainly 3d-character, are rather large, up to
23 eV for Fe. Within the 3d-like bands the
inter-band direct Coulomb matrix elements are of the same magnitude as the
intra-band matrix elements.  The matrix elements between 3d-states and
4sp-states are considerably smaller, of the magnitude of 13 - 14 eV. For
electrons in 4sp-states ($n,m \in \{5,\ldots,9\}$) the direct intra-band
Coulomb matrix elements are again of the order of 13 - 14 eV, but the
inter-band matrix elements are slightly smaller, about 9 eV.
The exchange matrix elements $J_{nm} = W_{nm,nm}$ are always
much smaller, usually less than 1 eV (for $n \neq m$).
The corresponding results for the other 3d-systems
investigated (Co, Ni and Cu) are very similar.\cite{Sch02}

% tab:UJtaverage
\begin{table}[!ht]
   \begin{tabular}{c|cccc}
    & $U$ & $J$ & $t_{NN}$ & $t_{NNN}$ \\
   \hline
   Fe & 21.1 & .81 & .59 & .24 \\
   Co & 22.6 & .87 & .55 & .10 \\
   Ni & 22.6 & .88 & .75 & .11 \\
   Cu & 24.5 & .94 & .80 & .12 \\
   \end{tabular}
   \caption{\label{tab:UJtaverage} Averaged on-site Coulomb, exchange,
     nearest neighbor and next nearest neighbor hopping matrix elements
     for the 4 3d-systems; energies are in eV.}
\end{table}

For the 5 states with predominant 3d-character we have calculated the averages
of the on-site direct and exchange Coulomb matrix elements
\begin{eqnarray} \label{eq:Fk_UJ}
   U &\equiv& \frac{1}{25} \sum_{mm'} W_{mm'm'm} \\
   J &\equiv& \frac{1}{20} \sum_{m\ne m'} W_{mm'mm'} ~~,
\end{eqnarray}
as well as the averages of the absolute values of the nearest neighbor (NN)
and next nearest neighbor (NNN) hopping matrix elements
\begin{equation}
   t_{NN(N)} \equiv \frac{1}{25} \sum_{n,m} |t_{\mR nm}| ~~.
\end{equation}
The results obtained thereby for the 4 transition metals under consideration
are shown in Table \ref{tab:UJtaverage}. The U-values vary between 21 eV for
Fe and 25 eV for Cu, the J-values are smaller than 1 eV and the hopping matrix
elements are of the magnitude 0.5 -- 0.7 eV for nearest-neighbor (NN) and
0.1 -- 0.2 eV for next nearest neighbor
(NNN), and further on decrease with increasing distance.

% tab:F^k
\begin{table}[!ht]
   \begin{tabular}{l|ccc}
    & $F^0$ & $F^2$ & $F^4$ \\
   \hline
   Fe (crystal)                      & 21.62 &  9.61 & 5.91 \\
   Fe (atom [\onlinecite{Watson59}]) & 23.76 & 10.96 & 6.81 \\
   \hline
   Co (crystal)                      & 23.18 & 10.31 & 6.34 \\
   Co (atom [\onlinecite{Watson59}]) & 25.15 & 11.58 & 7.20 \\
   \hline
   Ni (crystal)                      & 24.69 & 11.00 & 6.77 \\
   Ni (atom [\onlinecite{Watson59}]) & 26.53 & 12.20 & 7.58 \\
   \hline
   Cu (crystal)                      & 26.27 & 11.72 & 7.23 \\
   Cu (atom [\onlinecite{Watson59}]) & 27.90 & 12.82 & 7.96 \\
   \end{tabular}
   \caption{\label{tab:F^k} Slater integrals $F^k$ (in eV) for the 3d-systems
     Fe, Co, Ni, Cu as obtained by our calculations and within an earlier
     atomic calculation\cite{Watson59}.}
\end{table}

We have also evaluated the Slater integrals\cite{S29}:
\begin{equation} \label{eq:Fk_def}
   F^k \equiv e^2 \int dr~  r^2 \int dr'~ r'^2 ~
   |R_{l=2}(r)|^2 ~ \frac{r_<^k}{r_>^{k+1}} ~ |R_{l=2}(r')|^2 ~~,
\end{equation}
where $R_{l=2}(r)$ is a radial (atomic) d-wave function (obtained by
solving the Schr{\"o}dinger equation for a radial symmetric potential, for
instance). Note that only the three integrals $F^0$, $F^2$ and $F^4$
are required to determine all the Coulomb $d$-matrix elements.
Using the radial d-wave function obtained from the
Hartree calculation we obtain the following values for the Slater integrals
of the four 3d-systems: $F^0=$ 21.62 eV for Fe, 23.18 eV for
Co, 24.69 eV for Ni, and 26.27 eV for Cu. This means, the
Slater integrals $F^0$ are rather good estimates of our (averaged)
Coulomb matrix elements. These values are also in agreement with older results
obtained in calculations for 3d-atoms\cite{Watson59}. In Table \ref{tab:F^k} we
show our $F^k$-values for the four 3d-crystals and compare them with
corresponding atomic calculations from Ref. \onlinecite{Watson59}. Obviously,
there is fairly good agreement between these atomic and our results.

\section{\label{sec:HartreeFock}Unscreened Hartree-Fock approximation}

After we have determined the matrix elements within our restricted basis set
of 9 maximally localized Wannier functions (per site and spin), we have a
Hamiltonian in second quantization of the form
\begin{equation} % \label{eq:secondquant}
   H = \sum_{12\sigma} t_{12} c_{1\sigma}^{\dagger} c_{2\sigma} +
   \frac{1}{2} \sum_{1234\sigma\sigma'}
   W_{12,34} c_{1\sigma}^{\dagger}c_{2\sigma'}^{\dagger}c_{3\sigma'}c_{4\sigma}
\end{equation}
for which all the matrix elements are known from first principles. The
simplest approximation one can now apply is the HFA, which replaces the
many-body Hamiltonian by the effective one-particle Hamiltonian
\begin{eqnarray} \label{eq:HFHamilton}
   H_{\text{HF}} &=& \sum_{12\sigma} \left( t_{12}
                     + \Sigma_{12,\sigma}^{\text{HF}} \right)
   c_{1\sigma}^\dagger c_{2\sigma} \\
\mbox{with } ~~ \Sigma_{12,\sigma}^{\text{HF}} &=& \Sigma_{12}^{\text{Hart}}
   + \Sigma_{12,\sigma}^{\text{Fock}}
\\ \nonumber
   &=& \sum_{34\sigma'} \left[ W_{13,42} -
   \delta_{\sigma\sigma'}W_{31,42}\right] \bra c_{3\sigma'}^{\dagger}
   c_{4\sigma'} \ket ~~.
\end{eqnarray}
Here the expectation values $\bra c_{1\sigma}^{\dagger}c_{2\sigma} \ket$
have to be determined self-consistently for the HF Hamiltonian
(\ref{eq:HFHamilton}). Note that the Fock (exchange) term is
spin ($\sigma$) dependent and may, therefore, give rise to magnetic
solutions.

% fig1-4
\begin{figure}[!ht]
\includegraphics[scale=0.42]{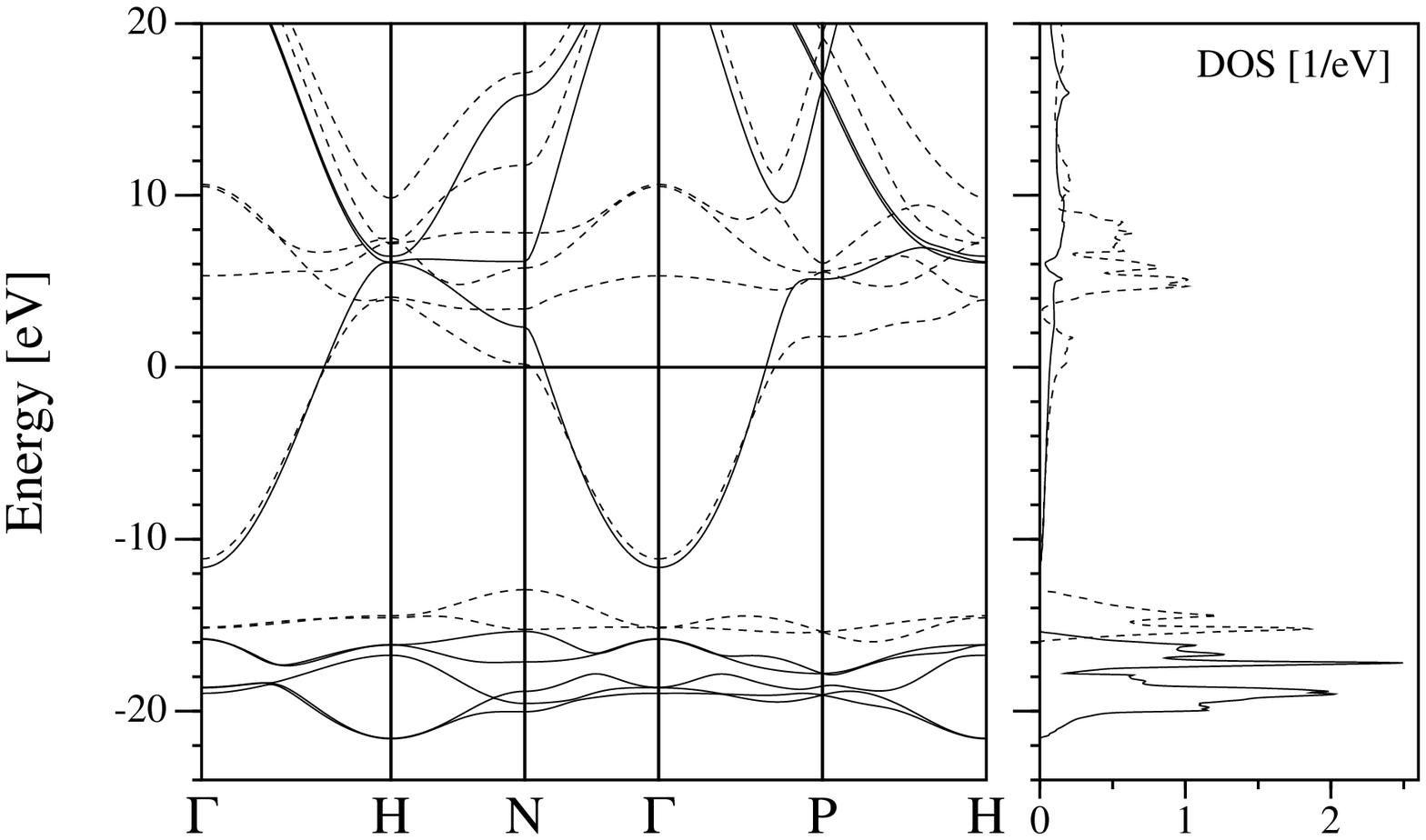}
\caption{\label{fig:fehf} Hartree-Fock band-structure and
   density of states (per spin) of Fe; the full line shows the majority (spin
   up), the dashed line the minority spin component.}
\end{figure}

\begin{figure}[!ht]
\includegraphics[scale=0.42]{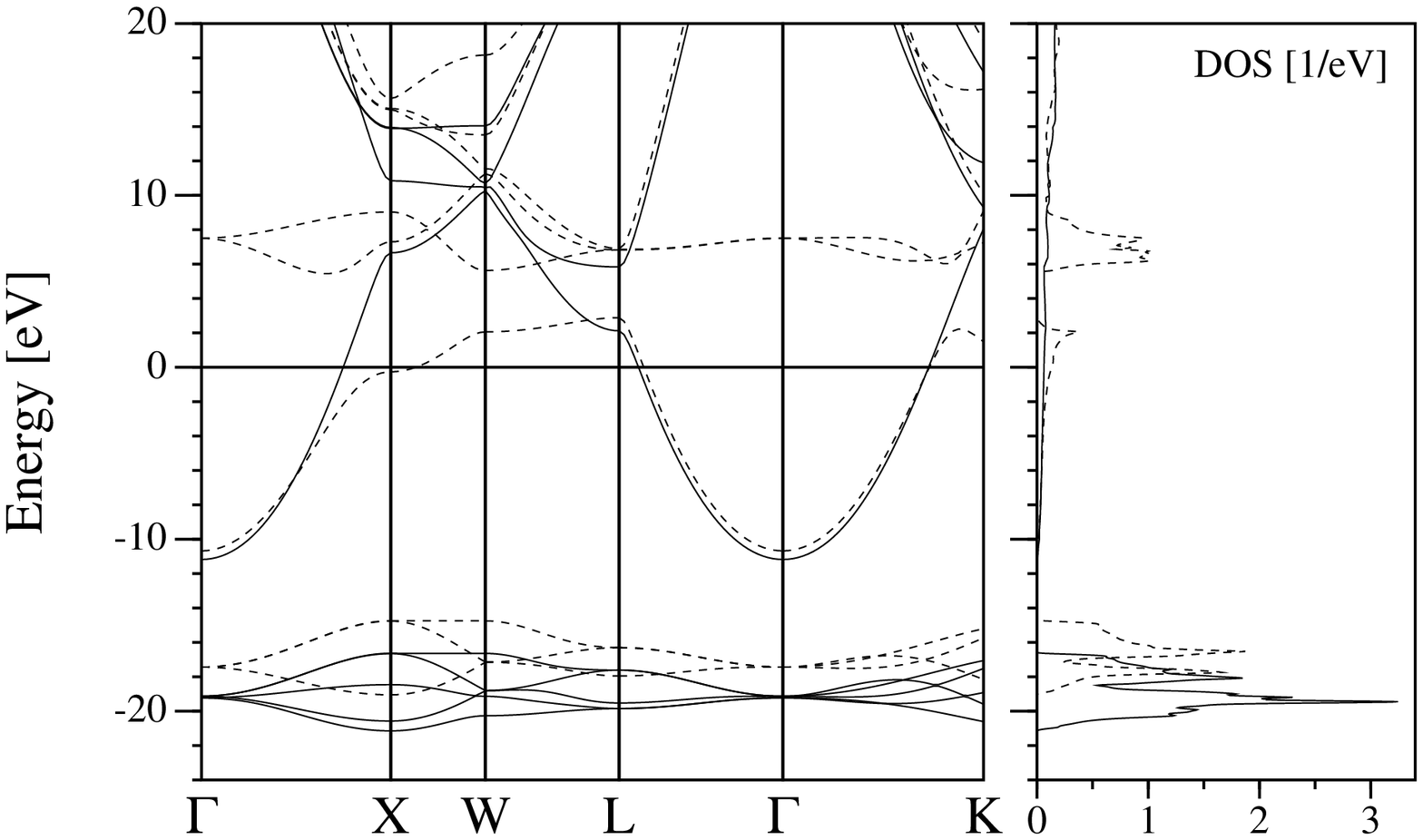}
\caption{\label{fig:cohf} Hartree-Fock band-structure and
   density of states (per spin) of Co; the full line shows the majority (spin
   up), the dashed line the minority spin component.}
\end{figure}

\begin{figure}[!ht]
\includegraphics[scale=0.42]{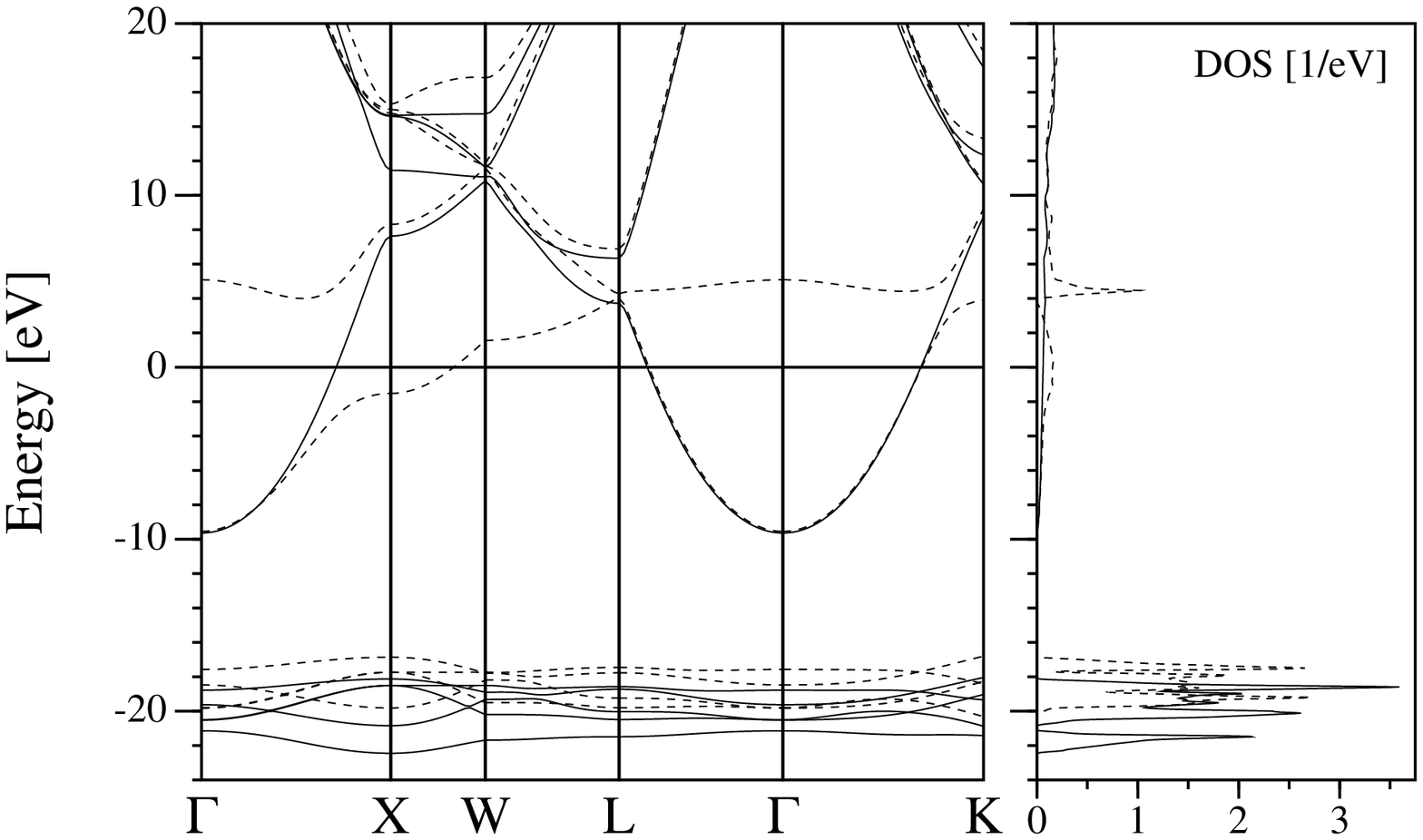}
\caption{\label{fig:nihf} Hartree-Fock band-structure and
   density of states (per spin) of Ni; the full line shows the majority (spin
   up), the dashed line the minority spin component.}
\end{figure}

\begin{figure}[!ht]
\includegraphics[scale=0.42]{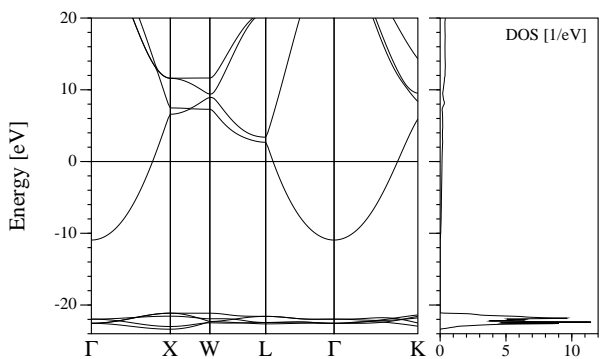}
\caption{\label{fig:cuhf} Hartree-Fock band-structure and
   density of states (for both degenerate spin directions) of Cu.}
\end{figure}

% tab:MM
\begin{table}[!ht]
   \begin{tabular}{c|cccc}
   ~~~~~ & ~~~HF~~~ & ~~EXX~~ & ~LSDA~ & Experiment \\
   \hline
   Fe & 2.90  & 3.27  & 2.18  & 2.22 \\
   Co & 1.90  & 2.29  & 1.58  & 1.72 \\
   Ni & 0.76  & 0.68  & 0.58  & 0.62 \\
   \end{tabular}
   \caption{\label{tab:MM} Spin magnetic moments ($\mu_B$/atom)
     from different methods.  The EXX results are from
     Ref. \onlinecite{Kotani98}.}
\end{table}

\begin{figure}[!ht]
\includegraphics[scale=0.42]{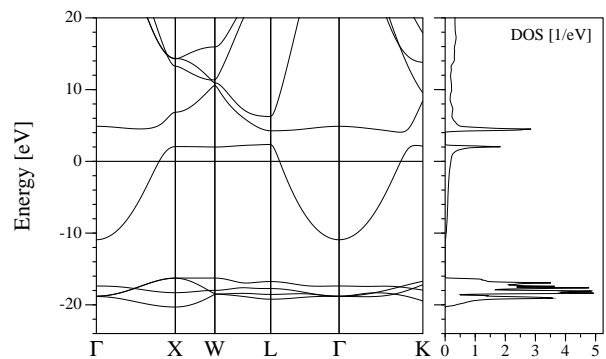}
\caption{\label{fig:conm} Non-magnetic Hartree-Fock band-structure and
   density of states (for both degenerate spin directions) of Co.}
\end{figure}

The Hartree-Fock results for the four materials of interest are shown in
Figs. \ref{fig:fehf}--\ref{fig:cuhf}. We show the effective
HF band structure and its density of states (DOS). In our HF calculations
there are no singularities (or a vanishing DOS) at the Fermi level since
we start from a localized description and consider
the Coulomb matrix elements only up to next neighbors.
Therefore, we implicitly truncate the Coulomb interaction in real
space and in practice work with an effective short-ranged interaction.
Within HFA the main part of the 3d-bands lies between 18 and 22 eV
below the Fermi level and is separated from the 4sp-bands. We find
magnetism in HFA for Fe, Co, and Ni in agreement with experiment. The
five majority spin d-bands are about 20 eV below the Fermi energy
and are completely filled. But the partially filled minority d-bands
have two (for Fe), three (for Co), and four (for Ni) filled bands between
-18 and -15 eV, and the rest are around and above the Fermi level.
The resulting magnetic moments are shown in Table~\ref{tab:MM}.
For copper no magnetism and exchange splitting of the 3d-bands is obtained,
but the (spin degenerate) 3d-bands are at about 22 eV below the Fermi level
and separated from the 4sp-bands. If we compare these results with the
results of the simple Hartree approximation, which are qualitatively similar
to LDA results (as shown e.g. in Ref. \onlinecite{MJW78},
or in our detailed results\cite{Sch02}),
we see that the exchange term has two effects:
It produces an exchange splitting and the possibility of magnetic
solutions, and it draws the 3d-bands energetically down by an amount of
about 20 eV.
Compared with experiment the HFA overestimates magnetism and leads to overly
large values for the magnetic moment, see Table~\ref{tab:MM}.
This is consistent with Heisenberg or Ising model studies where the
mean-field approximation HFA also has the tendency to overestimate
magnetism and magnetic solutions.
However, the reason why the 3d-bands lie so far below the Fermi level and the
4sp-band in HFA has nothing to do with the existence and overestimation of
magnetism. This can be seen already from the non-magnetic system Cu, for
which the (fully occupied) 3d-bands also lie at about 22 eV below
the Fermi level (see Fig. \ref{fig:cuhf}).
To demonstrate this also for a system with a partially filled 3d-band we have
done a non-magnetic Hartree-Fock calculation for Co (by forcing
equal occupation for both spin directions).  The results for the band structure
and the DOS are shown in Fig. \ref{fig:conm}.
We observe again that the main part of the 3d-bands are well below the
4s-bands and Fermi level; note the hybridization gap
caused by the unoccupied 3d-bands above the Fermi level.

% fig5

\section{\label{sec:atomicHFA} Comparison with atomic Hartree-Fock results}

We have seen in the previous section that one effect of the HFA calculation,
when compared with the Hartree calculation, is the shift of the 3d-bands
down (about 20 eV below the Fermi level and about 8--10 eV below the
bottom of the 4sp-band). This shift of the d-bands is about the same energy as
the Coulomb matrix elements $U$, and roughly agrees with earlier atomic
Hartree-Fock calculations\cite{Watson59,HodgesWatson72},
where the 3d-states are also about 10 eV below the 4s-states.

% fig6
\begin{figure}[!ht]
\includegraphics[scale=0.45]{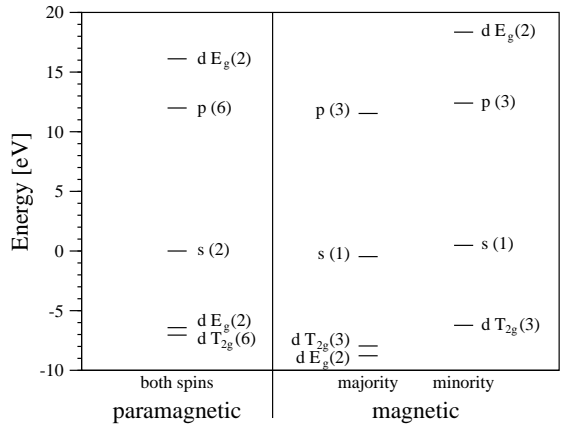}
\caption{\label{fig:coatom} Energy eigenvalues from quasi-atomic
   HFA-calculation.  The numbers in brackets indicate the degeneracy.}
\end{figure}

Because the inter-site hopping matrix elements in Table \ref{tab:UJtaverage}
are much smaller than the U-values one may consider an expansion in
$t/U$, with the zeroth order approximation to completely neglect hopping.
Doing this, we have performed a quasi-atomic HFA calculation for Co,
by including only the on-site one-particle and two-particle (Coulomb)
matrix elements. The results are summarized in Fig. \ref{fig:coatom};
the degeneracy of the different levels is also indicated. In the paramagnetic
case,we find that the 3d-bands are below the 4s-bands (at the Fermi level) by
about 6 to 7 eV, which is in rough agreement with the earlier atomic
HFA results\cite{Watson59,HodgesWatson72}. The splitting between the
occupied and unoccupied 3d-states is about 23 eV, which is the on-site
$U$ for Co. Magnetic HFA solutions are also found in the atomic
limit for Co, as shown in the right panel of Fig. \ref{fig:coatom}.
The majority-spin 3d states ($T_{2g}$ and $E_g$) are now completely filled
and energetically lie lower than the corresponding non-magnetic
HFA-states. But only the (3-fold degenerate) $T_{2g}$-states of the
minority-spin electrons are filled whereas the $E_g$-states of the minority
electrons are empty (and now even 26 eV above the occupied d-states).
The additional energetical shifts between the occupied 3d-states in the
paramagnetic and ferromagnetic atomic HFA solution are due to the exchange
matrix elements $J$.

This behavior can qualitatively be understood within the framework of the
following simple, analytically solvable model.  Similar to
the numerical HFA-results presented and discussed above, we neglect
all inter-site one-particle (hopping) and interaction matrix elements.
Furthermore, we assume that we have diagonalized the one-particle Hamiltonian,
taking into account only the atomic
3d-levels and assuming that the on-site one-particle diagonal matrix elements
$\varepsilon$, the Coulomb matrix elements $U$, and the
exchange matrix elements $J$ are equal, i.e., that the 3d-levels are
degenerate in the atomic limit with no crystal-field effects. Then the
atomic part of the many-body Hamiltonian can be
written as
\begin{eqnarray} \nonumber
   H = \sum_{i\sigma} \varepsilon ~ c_{i\sigma}^{\dagger} c_{i\sigma} &+&
   \frac{U}{2} \sum_{(i\sigma) \neq (j\sigma')}
     c_{i\sigma}^{\dagger} c_{i\sigma} c_{j\sigma'}^{\dagger} c_{j\sigma'}
\\
   &+& \frac{J}{2} \sum_{i \neq j, \sigma \sigma'}
     c_{i\sigma}^{\dagger} c_{j\sigma'}^{\dagger} c_{i\sigma'} c_{j\sigma}
\end{eqnarray}
where $i,j \in \{0,\ldots,4\}$ denote the 5 (degenerate) 3d-states.

The standard Hartree-Fock decoupling leads to
\begin{eqnarray} \nonumber
   H = \sum_{i\sigma} \Big( \varepsilon &+& U \big[ \sum_{j\sigma'}
   \langle c_{j\sigma'}^{\dagger} c_{j\sigma'} \rangle
   - \langle c_{i\sigma}^{\dagger} c_{i\sigma} \rangle \big]
\\
   &-& J \sum_{j \neq i} \langle c_{j\sigma}^{\dagger} c_{j\sigma} \rangle
   \Big)
   c_{i\sigma}^{\dagger} c_{i\sigma} ~~.
\end{eqnarray}
Here we have assumed that the Hartree-Fock Hamiltonian has the same symmetry
as the uncorrelated Hamiltonian, and hence off-diagonal expectation values
$\langle c_{j\sigma'}^{\dagger} c_{i\sigma}\rangle$ for
$(i\sigma) \neq (j\sigma')$ vanish.
 From this equation it is clear that the HF Hamiltonian can be written
in terms of an effective one-particle energy
\begin{equation}
  H = \sum_{i\sigma} \varepsilon_{i\sigma}^{\text{HFA}}
        c_{i\sigma}^{\dagger} c_{i\sigma}
\end{equation}
where
\begin{equation}
   \varepsilon_{i\sigma}^{\text{HFA}} =
   \varepsilon + U \big[
     \sum_{j\sigma'}
       \langle c_{j\sigma'}^{\dagger} c_{j\sigma'} \rangle
     - \langle c_{i\sigma}^{\dagger} c_{i\sigma} \rangle
   \big]
   - J \sum_{j \neq i} \langle c_{j\sigma}^{\dagger} c_{j\sigma} \rangle .
\end{equation}
In the simple Hartree approximation (HA) the exchange decouplings are
neglected, which means that all the decoupling terms with the negative
sign would not occur. Therefore, the corresponding Hartree one-particle
energies are given by
\begin{equation}
   \varepsilon_{i\sigma}^{\text{HA}} =
   \varepsilon + U \sum_{j\sigma'}
   \langle c_{j\sigma'}^{\dagger} c_{j\sigma'} \rangle ~~.
\end{equation}
Comparing this result with the Hartree-Fock one-particle energies, we find
that the HF occupied levels are shifted downwards by an amount of
\begin{equation}
   U \langle c_{i\sigma}^{\dagger} c_{i\sigma} \rangle +
   J \sum_{j \neq i} \langle c_{j\sigma}^{\dagger} c_{j\sigma} \rangle
\end{equation}
relative to the Hartree levels.
Momentarily setting $J=0$, we see that for $N$
occupied levels the Hartree approximation gives the one-particle energies
\begin{equation}
   \varepsilon_{i\sigma}^{\text{HA}} = \varepsilon + N U
\end{equation}
whereas the HFA yields
\begin{equation}
   \varepsilon_{i\sigma}^{\text{HFA}} = \varepsilon + (N-1) U ~~.
\end{equation}
The occupied Hartree-Fock one-particle energies are lower than the
corresponding Hartree one-particle energies by $U$, which
is a consequence of the artificial and unphysical self-interaction still
present in the Hartree approximation that is exactly canceled in
Hartree-Fock. This also explains why the
Hartree-Fock bands are shifted downwards from the Hartree bands by
an energy of the amount $U$. One also sees from this simple atomic-limit
Hartree-Fock model that the energy difference between the highest
occupied and the lowest unoccupied effective Hartree-Fock one-particle
energies is again essentially $U$, which is once more in agreement with our
numerical results for the crystal and for the atom (cf. Fig.\ref{fig:coatom}).
Note that we have ignored $U_{sd}$ interactions, which cause an additional
shift of d-bands below the s-bands by about an additional 10 eV in
the full HFA calculations.

Taking into account the exchange interaction $J$ again and denoting by
$N_{\sigma}$ the number of occupied states with
spin $\sigma$ (i.e. $N = N_{\uparrow} + N_{\downarrow}$) one obtains in HFA
\begin{equation}
   \varepsilon_{\sigma}^{\text{HFA}} =
   \varepsilon + (N-1) U - (N_{\sigma}-1) J ~~.
\end{equation}
Then the total energy in HFA is given by
\begin{equation}
   E_{\text{tot}} = N \varepsilon + \frac{N(N-1)}{2} U -
   \sum_{\sigma} \frac{N_{\sigma}(N_{\sigma}-1)}{2} J ~~.
\end{equation}
For the total energy we have added the necessary correction term to the
sum of the occupied energy levels (much like the double counting
term that shows up in band-structure calculations).
Now for partially filled 3d-shells the occupation of the different spin
directions may be different. Denoting $M = N_{\uparrow} - N_{\downarrow}$ we
obtain for the total energy
\begin{equation}
   E_{\text{tot}} = N \varepsilon + \frac{N(N-1)}{2} U - \frac{N^2+M^2}{4} J
   + \frac{N}{2} J ~~.
\end{equation}
The magnetic ($M \neq 0$) total energy is lower than the non-magnetic
(consistent with Hund's rules).

Take once more Co with 8 3d-electrons.  The paramagnetic
(non-magnetic) state has the occupations
$N_{\downarrow} = N_{\uparrow} = 4$ ($N$=8 and $M$=0).
For this configuration (corresponding to the left panel in
Fig. \ref{fig:coatom}) one obtains
\begin{equation}
   E_{\text{tot}}^{(P)} = 8 \varepsilon + 28 U - 12 J ~~.
\end{equation}
The Hund's rule magnetic solution has 3d-states of
one spin-direction completely filled, i.e.,
$N_{\uparrow} = 5$ and $N_{\downarrow} = 3$
($N$=8 and $M$=2).
This gives
\begin{equation}
   E_{\text{tot}}^{(M)} = 8 \varepsilon + 28 U - 13 J ~~.
\end{equation}
Therefore, the magnetic configuration (with a magnetic
moment of 2 for the atom) is energetically more favorable by $J$.
Note also the exchange splitting in the occupied energy eigenvalues
\begin{equation}
   \varepsilon_{\downarrow} - \varepsilon_{\uparrow} = 2J
\end{equation}
and that our model would predict the unoccupied minority spin $E_{2g}$ state
to be $U+J$ higher in energy than the corresponding occupied majority spin
state.

The simple model in this section differs from the results shown in
Fig.~\ref{fig:coatom} in that we have replaced the full matrix of
$U$ and $J$ by scalar values for d-states only
(ignoring s-d interactions, for example).
However, it captures all of the important physics without attempting to
be completely quantitative.

\section{\label{sec:LSDA} Comparison with LSDA and EXX results}

For comparison with the HFA results described in Section
\ref{sec:HartreeFock} we have also performed a standard
LSDA band-structure calculation with the LMTO-ASA method.
We used the von Barth-Hedin exchange-correlation potential\cite{BH72}.
Since these results are very similar to those of
Ref. \onlinecite{MJW78}, we do not repeat them here.
Again, our detailed results are given in Ref. \onlinecite{Sch02}.
For the magnetic systems Fe, Co, and Ni
we obtain an exchange splitting and the prediction of magnetic
solutions with magnetic moments shown in Table~\ref{tab:MM},
which are in better agreement with experiment than the HFA results.

The energy spectra of the bands (DOS) are quite different from the HFA.
For example, the 3d-bands now fall into the same energy region as
the 4sp-bands, i.e., the LSDA-results are not so different from the
Hartree-results. This means that the exchange-correlation energy leads only
to a small shift of the 3d-bands downwards by at most a few eV and a smaller
exchange splitting (also of the magnitude of 1 eV).  
On the other hand, in the LSDA
calculations the self-interaction terms are not completely canceled, i.e.,
an (unrealistic) self-interaction is included, which may lead to 3d bands
that lie energetically too high, as discussed for the atomic limit in the
previous section.

To see the effect of correlations within LSDA, we have also
performed an exchange-only calculation for Co, i.e. only the exchange
part of the (local) exchange-correlation potential\cite{BH72} was employed.
The result is similar to the LSDA result,
and the (majority) d-bands lie only about 1~eV lower than within LSDA,
i.e., very minimal when compared with the large drop in the full HFA.
This exchange-only LSDA-result also contains self-interactions,
and their exact cancelation in the HFA is responsible for the large shift
downwards of the d-bands.
Nevertheless, the LSDA result indicates that a possible
effect of correlations is to shift the 3d-bands up relative to exchange-only
calculations, and hence one would expect a similar effect if correlations
could be added to the full HFA calculations.

% fig7
\begin{figure}[!ht]
\includegraphics[scale=0.45]{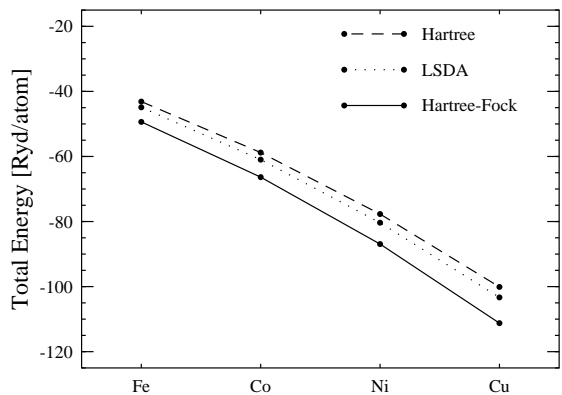}
\caption{\label{fig:etot} Total ground state energy (of the valence electrons)
   obtained in Hartree-approximation, LSDA and HFA for the 3d transition
   metals Fe through Cu.}
\end{figure}

One can also calculate the total energy in the Hartree, HFA, and LSDA
approximations. The results obtained for the four materials of interest are
shown in Fig. \ref{fig:etot}. We see that the total energy is always
significantly
lower in HFA than in the Hartree approximation, which is expected because
the HFA minimizes the total energy.  The HFA total energy is also
lower than the L(S)DA, and the LSDA result is lower than the
simple Hartree result.  Because of the unknown approximations that go
into constructing L(S)DA, it is hard to guess ahead of time that this would be
the case.  However, it is well known that the L(S)DA approximation produces a
bad total exchange-correlation energy; the reason why such good agreement
with experiment is found is that relative exchange-correlation energies are
nonetheless reasonably accurately calculated.

We now turn to the comparison of our results with exact exchange (EXX)
calculations for 3d systems\cite{Kotani97,Kotani98}.
This method, which is self-interaction-free, uses the EXX energy\cite{Kotani97}
instead of a LDA local exchange and then adds in a local LDA correlation
potential.
Like the HFA the magnetic moments for Fe, Co and Ni (see Table~\ref{tab:MM})
are overestimated by EXX\cite{Kotani98}.
Our HFA results for Fe show the majority 3d-bands about 20 eV below
the Fermi level (Fig.~\ref{fig:fehf}), whereas the EXX density of states
(Fig.~2 in Ref.~\onlinecite{Kotani97}) show these bands only about 10 eV below
the Fermi level.  The differences are probably due to the (LDA) correlations
shifting the 3d-bands upwards.
This upward shift also occurs with EXX+RPA results in
Ref.~\onlinecite{Kotani98}.  Here the LDA correlations present in EXX are
replaced by RPA correlations and the 3d-bands are found in the region of the
4sp-bands (similar to LDA).
It is likely that the qualitative agreement between the HFA and EXX
results for Fe is due to the correct cancelation of self-interactions.

\section{\label{discussion} Discussion and conclusion}

We have presented the results of (unscreened) HFA calculations
for the 3d transition metals Fe, Co, Ni and Cu.
We obtain magnetic solutions for Fe, Co, and Ni with (slightly) too large
magnetic moments when compared to experimental or LSDA results. The occupied
HFA 3d-bands lie about 20 eV below the Fermi level (and the Hartree result),
which is also the magnitude of the splitting between occupied and unoccupied
3d-bands and of the magnitude of the on-site Coulomb matrix element (the
``Hubbard'' U). This downwards shift of the HFA 3d-bands compared to the
Hartree- and LSDA-3d-bands can be understood as due to the self-interaction
correction of HFA.

One may argue that these results are not surprising and an artifact of
using the unscreened HFA.  Our ab-initio calculation of
the direct Coulomb matrix elements yields large values of the magnitude of
20 eV.  HFA can be considered to be an approximation for the selfenergy which
is correct only in linear order in the Coulomb interaction.
But for these large values of the U-terms HFA is certainly not sufficient but
one has to apply better many-body approximations.
One should apply systematic extensions of HFA,
which within the standard perturbational approach can be
represented by (a resummation of an infinite series of) Feynman diagrams, or
one can try to apply the recently so successful non-perturbational many-body
schemes like ``dynamical mean field theory'' (DMFT)\cite{GKKR96} or
variational (Gutzwiller) approaches\cite{GebhardWeber}.
The simplest standard diagram series
are the bubble diagrams leading essentially to the ''random phase
approximation'' (RPA). This means just a renormalization of the interaction
line, i.e. the pure ``naked'' Coulomb interaction has to be replaced by a
``dressed'' interaction. Or in other words, the exchange (Fock) contribution
has not to be calculated with the bare Coulomb matrix elements but with
screened Coulomb matrix elements. Probably the non-perturbational schemes
like DMFT are also only applicable for screened Coulomb matrix elements.

We believe that the approach we have used for our HFA calculations
can easily be generalized to provide an approach for combining ab-initio
and many-body methods for the calculation of the electronic properties of
solids. The starting point is a traditional band-structure calculation
for an effective (auxiliary) one-particle Hamiltonian, which can be the
Hartree-Hamiltonian.  This yields, in particular, the eigenfunctions in
the form of Bloch functions.  Keeping only a finite number of $J$ band
indices restricts and truncates the Hilbert space for further
calculations.  We use the Marzari-Vanderbilt algorithm to construct
maximally localized Wannier functions (within the truncated one-particle
Hilbert space).  All the one-particle (tight-binding) and
two-particle (Coulomb) matrix elements between these Wannier functions
can be calculated. The strong localization guarantees that only on-site
matrix elements and near-neighbor inter-site matrix elements have to be
calculated.  We are left with a many-body Hamiltonian in second
quantization but with parameters determined from first principles for
any given material, which we have solved within the HFA but for which 
we should also be able to solve by using more sophisticated many-body 
techniques.
Our HFA approach is free from the problems of double counting of
correlation effects and self-interaction and considers exchange
contributions exactly.  It does not rely on assumptions based
on the homogeneous electron gas or a dependence on the
local electron density.  An inhomogeneous (lattice) electron system is
considered right from the beginning.
Within the standard Feynman diagram approach
the most straightforward next step beyond HFA would be a summation of
bubble diagrams leading to a renormalized (screened) Coulomb interaction.
This would require calculating the exchange contribution not with the 
bare but with
a screened Coulomb interaction.  To take into account the effects of screening
would require a calculation of the charge susceptibility and the 
(static) dielectric constant, which could be done within a generalized
Lindhard theory, for instance.

\begin{acknowledgments}
   This work has been supported by a grant from the Deutsche
   Forschungsgemeinschaft No. Cz/31-12-1.
   It was also partially supported by the Department of Energy
   under contract W-7405-ENG-36.  This research used resources of the
   National Energy Research Scientific Computing Center, which is
   supported by the Office of Science of the U.S. Department of Energy
   under Contract No. DE-AC03-76SF00098.
\end{acknowledgments}

\end{document}